\documentclass[12pt,preprint]{aastex}

\begin{document}
%
%
\input epsf
\def\refpar{\par\hangindent=3em\hangafter=1}
\def\endselectedbib{\refpar\egroup}
\def\pageassign#1#2{[pages: #1, author: #2]}

\title{Stability of the Submillimeter Brightness of the Atmosphere Above
Mauna Kea, Chajnantor and the South Pole}

\author{J.~B.~PETERSON\altaffilmark{1},
S.~J.~E.~RADFORD\altaffilmark{2},
P.~A.~R.~ADE\altaffilmark{3},
R.~A.~CHAMBERLIN\altaffilmark{4},
M.~J.~O'KELLY\altaffilmark{1}\altaffilmark{5},
K. ~M. ~PETERSON\altaffilmark{1} and
E. SCHARTMAN \altaffilmark{1}\altaffilmark{2}\altaffilmark{6}\\}

\altaffiltext{1} {Department of Physics, Carnegie Mellon University, 5000 Forbes Ave. Pittsburgh PA
15213, email: jbp@cmu.edu \\}
\altaffiltext{2} {National Radio Astronomy Observatory, 949 North Cherry Ave., Tuscon AZ 85721-0665, 
email: sradford@heineken.tuc.nrao.edu\\}
\altaffiltext{3} {Physics and Astronomy Department, Cardiff University, CF2 3YB Cardiff, UK\\}
\altaffiltext{4} {California Institute of Technology, 
Submillimeter Observatory, 11 Nowello St, Hilo HI 96720\\}
\altaffiltext{5} {Current address: Department. of Physics, Massachusetts Institute of Technology,
Cambridge, MA 02139\\} 
\altaffiltext{6} {Current address: Princeton Plasma Physics Laboratory, 
Princeton University, Princeton NJ 08540 \\}


\begin{abstract}

{\baselineskip 1.2pc

The summit of Mauna Kea in Hawaii, the area near Cerro Chajnantor 
in Chile, and the South Pole are sites of large millimeter or 
submillimeter wavelength telescopes.  
We have placed 860 GHz sky brightness monitors at all three sites and
present a comparative study of the measured submillimeter brightness due to atmospheric 
thermal emission. We report the stability of that quantity at each site.  
}

\end{abstract}


\section{ Introduction }

Observations at submillimeter wavelengths are used to study molecular clouds, 
the Galactic center region, 
and high redshift galaxies, among other sources. 
At submillimeter wavelengths these astrophysical sources are more transparent than
at optical or infrared wavelengths, allowing the study of the interiors of dense regions.
However, despite the transparency of astronomical sources, 
the Earth's atmosphere is largely opaque at these
wavelengths.

Water vapor and other atmospheric gases absorb submillimeter radiation, 
attenuating astronomical signals. By Kirchoff's law, this absorption also leads to thermal 
radiation from the atmosphere itself, adding
noise. Together, these effects degrade the signal to noise ratio, 
increasing the observing time necessary to reach a desired sensitivity. 
Worse, moving inhomogeneities in the water vapor distribution cause rapid fluctuations in the
atmospheric radiation, called sky noise, and slow variations of attenuation make calibration
difficult. The uncertainty in astronomical flux caused by fluctuations due to the atmosphere
often greatly exceeds that due to receiver noise.

Because of sky noise and attenuation, submillimeter
astronomical observations can only be made at the driest, most
stable sites.   Three premier sites that have been used for millimeter
or submillimeter observations and have also been chosen for large
submillimeter telescopes are Mauna Kea in Hawaii (4100 m elevation), the area near
Cerro Chajnantor in Chile (5000 m), and the Amundsen-Scott
Station at the South Pole (2800 m).  We placed
identical tipping photometers at these three sites to make automated
measurements of sky brightness at 860 GHz (350 micrometers wavelength).
These data are presented here in the form of values of the zenith optical depth of the 
atmosphere.

Previous measurements have shown that all
three sites have very low {\it millimeter} wavelength (30-300 GHz) sky brightness. Identical
radiometers operating at 225 GHz have been used for years to measure the
brightness of the atmosphere. The median zenith
optical depths derived from these measurement are: 
Mauna Kea--0.091, Chajnantor--0.061, South Pole--0.053. (Chamberlin \& Bally 1994, 
Chamberlin, Lane \& Stark 1997, Radford \& Chamberlin 2000) 
Data from these instruments have not been analyzed to compare 
the atmospheric stability at the sites.

One previous study compared estimates of
centimeter and millimeter wavelength sky noise at the South
Pole and Chajnantor (Lay \& Halverson 1999). A 11.2 GHz phase difference monitor has been
operated at Chajnantor for several years (Radford, Reiland, \& Shillue, 1996).  
This interferometer has a baseline
of 300 m and measures phase fluctuations in a carrier wave
broadcast by a geostationary satellite. These phase
fluctuations are primarily due to turbulence in the water vapor above
the site.  At the South Pole the Python telescope was used to
study the cosmic microwave background at 90 GHz.
The atmospheric emission fluctuations
seen in the Python data are also due to water vapor variations.  
These data can be compared, but
because the two instruments measure different spatial scales and different
frequencies, models of 
atmospheric turbulence and the
atmospheric emissivity spectrum are needed for comparison.
In addition, to compare the two sites, an assumption must be made about the
height of the turbulent layer at each site. 
Making reasonable assumptions, the analysis of Lay and Halverson indicates the 
sky noise at South Pole is lower than at Chajnantor by about a factor 10.

Narrow band submillimeter transparency measurements have been made for several years
using the AST/RO telescope at the South Pole (Chamberlin, Lane \& Stark 1997) at 461 and 492 GHz.
This region of the spectrum is complicated by the wings of nearby O$_2$ and H$_2$O lines.
The median winter zenith optical was found to be 0.70.
Opacity at the South Pole has also been measured at 806 GHz with AST/RO and from 100 to 1800 GHz
using a Fourier Transform Spectrometer (Chamberlin, Martin, 
Martin \& Stark 2002).
The South Pole data have not been analyzed to study stability and no similar data
exist for Mauna Kea or Chajnantor. 

To provide a set of site quality measurements in a standard band used for submillimeter observations 
we built a set of identical
site monitor instruments, called submillimeter tippers, which measure 
the sky brightness in the 860 GHz spectral window. 
These tippers overcome two of the problems of some of the previous site evaluation 
observations, dissimilar instrumentation and sporadic observation.
The tippers are designed to be simple and compact, require little electrical power, and no cryogens.  
The tippers have the sensitivity to measure the zenith 
optical depth to about three percent every 12 minutes. 

The 860 GHz atmospheric window lies between two strong pressure broadened water lines. 
Because of the
proximity to water lines, for most of the data presented here, water vapor is the component of the
atmosphere that contributes most to opacity.
At the South Pole, however, there is evidence of a dry air component 
of opacity. After applying the window correction of Calisse (2002) to the data
presented in Chamberlin (2001) we find $\tau_{dry} \sim 0.3 \pm 0.1$.  Since $\tau$
values as low as 0.5 are measured at all three sites the dry air component
is significant at times.

Broadband submillimeter astronomical observations
are usually made by beam switching that is, by rapidly (1-10 Hz) moving the telescope beam between
points on the sky.  By differencing the measured flux from two or more nearby points
much of the atmospheric emission can be subtracted from the data. It is the residual atmospheric
emission after this subtraction that acts as the sky noise for these observations.
Our submillimeter tipper instruments are not sensitive 
enough to detect variations of sky brightness at 10 Hz, and the beam width
of our instrument (6$^\circ$) is much larger than that of
a large submillimeter telescope. To use the variations of optical depth
reported below as an indication of sky noise in beam switched observations would require an
extrapolation by about $10^3$ in both angular scale and fluctuation frequency. 
The uncertainty in that extrapolation exceeds the site to site differences in sky noise presented here.

To calibrate submillimeter observations, measurements of the atmospheric attenuation are needed
throughout the course of the observation.  The data presented here are taken in a manner and over a time scale
similar to the attenuation measurements typically made during submillimeter observations.  
When interpreted as measures of attenuation stability, these results
apply directly, with no need for extrapolation.  

In this paper we present data from all three sites for the period January 1998 to August 2002.

\section{Instrument Description}

Radiation from the sky enters the tipper through a cylindrical
window of woven expanded PTFE then is focused and 
redirected by an off-axis parabolic mirror.  The sky flux is periodically 
interrupted at 0.75 Hz by a rotating chopper wheel.  
After the chopper, the radiation is filtered 
by a multilayer resonant mesh filter and then enters a compound parabolic concentrator (CPC)
that illuminates a pyroelectric detector. The entrance aperture diameter of the CPC
and the focal length of the parabolic mirror together determine the 6$^\circ$ beam 
width of the tipper.

The parabolic mirror is mounted on the shaft of a stepping motor so  
the mirror can be rotated about the axis of the CPC, allowing observations 
at a various elevation angles.  Rotating the
mirror also allows the beam to be directed into either of two blackbody 
calibrators, which are held at approximately 300K and
340K.

Approximately five times each hour a set of observations, which we call a tip, are 
made.  A tip
consists of observations at elevation angles $\theta_i$= 90$^\circ$, 41.8$^\circ$, 30$^\circ$, 23.6$^\circ$,
 19.5$^\circ$, 16.6$^\circ$ and 14.5$^\circ$ 
(chosen to provide a linear progression of airmass) along with observations 
of both calibrators.  

Further details on the design and operation of the instrument are presented in 
Radford, Peterson, Schartman, Valentine \& Chamberlin, (2002).  

\section{Data Reduction}

In the data reduction brightness data from the tips are converted
to values of slab-model zenith optical depth $\tau$. 
This optical depth determines both the emission from and
attenuation by the atmosphere.

First,  calibration measurements are used
to convert the measured detector voltages to sky brightness temperatures. 
Then, data for each tip are least-squares fit to an isothermal slab model of the
atmosphere, 
$T_i=T \left( 1-e^{-\tau \cdot A_i} \right) $.
Here $\tau$ is the zenith optical depth, $T$ is the effective temperature 
of the atmosphere, $A_i={\rm csc}(\theta_i)$ are airmass values, $T_i$ are  
sky brightness temperatures, and  the index $i$ counts through the
seven elevation angles used in the tip. 
In addition to estimating the optical depth of the atmosphere and its effective 
temperature, the processing provides uncertainties in both parameters as well 
as the mean squared error and goodness of fit.  

The window of the tipper has an average transmittance $0.81 \pm 0.03$ across the passband of the
instrument, and the calibrators are inside the window. The window attenuates 
sky brightness signals and adds a window-emission brightness component.
We have corrected the $\tau$ values to remove the effect of window emission and attenuation
following the procedure of Calisse (2002).  Typical
values of the correction are are $-0.25 $ at $\tau = 2$ and $-0.32 $ at $\tau = 0.5$.
To check this window correction we temporarily added a second layer of window material to the
South Pole tipper.  Averaging 5 trials of this experiment we find the apparent 
value of $\tau$ was increased from 0.9 by 0.18 $\pm$ 0.07, consistent with the calculated
correction.   Note that all previous publications that make use of tipper results have not been
corrected for window emission and should be updated.

During 1999 instability of the chopping wheel speed in the South Pole tipper
caused increased and variable instrument noise. The variance values for the Pole in 1999
have been cut from the data set of variations of $\tau$.
The average values were not affected by the instability
so those values are included.

Some data fail to fit the slab model, either because of cloud structure in the
atmosphere or because of instrument malfunction.
An observation was discarded if
the best fit value of $\tau$ was greater than fifty or equal to zero,  
the uncertainty in $\tau$ was greater than 1000 or 0,
the estimated atmospheric temperature $T_{atm}$ was greater than 1000 K,
or the goodness of fit was 0.  Samples with fit values of $T_{atm}$ between 300K and
1000K are also suspect, since the real atmospheric temperature is well below
300K, but these samples were retained to avoid the possibility of biased removal of
high noise samples. Altogether, these data filters removed about 5\% of the data.

The data filters
might remove some legitimate data points with greater than average values 
of $\tau$ or $\sigma_{\tau}$. For submillimeter 
astronomy, however, only the stable periods with low $\tau$ are usable.
In any comparison of quiet sky periods the bias resulting from the filtering process
should be small and slightly favor a less stable site.  
Considering the small fraction
of data cut, the looseness of the cut parameters, and the
tendency of the bias to affect mostly the high $\tau$ segments 
of the data, we estimate systematic bias from data cutting
is less than one percent and make no correction for this effect.

\section{Qualitative Comparison}

\begin{figure*}
{\centering\leavevmode\epsfysize=3.75in 
\epsfbox{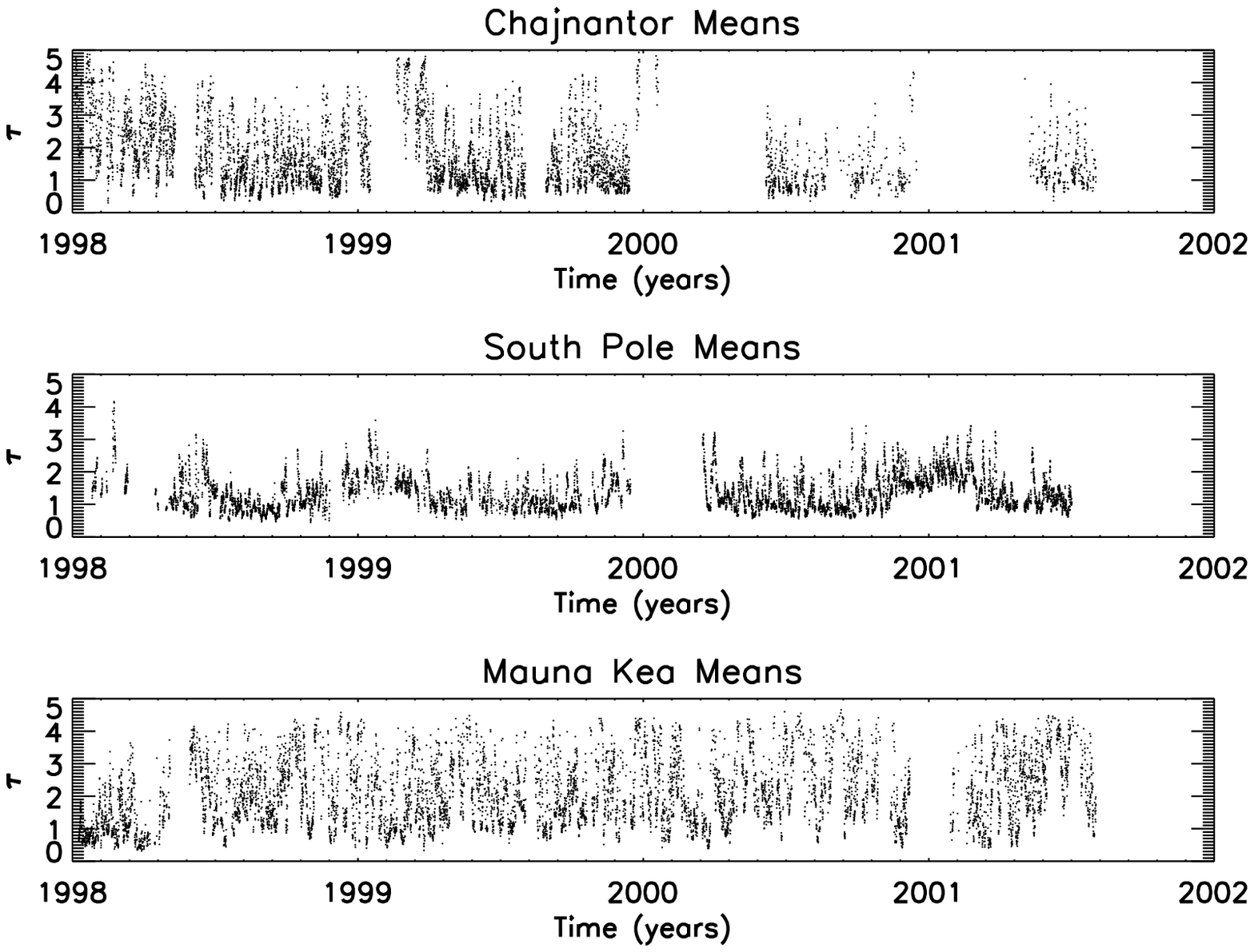}
}
\figcaption{\label{allyears}{\bf Zenith Optical Depth at 860 GHz. } 
Mean values for two hour intervals are plotted. Intervals with no data are
the result of tipper breakdowns or data cuts.   }
\end{figure*}

\begin{figure*}
{\centering\leavevmode\epsfysize=3.75in 
\epsfbox{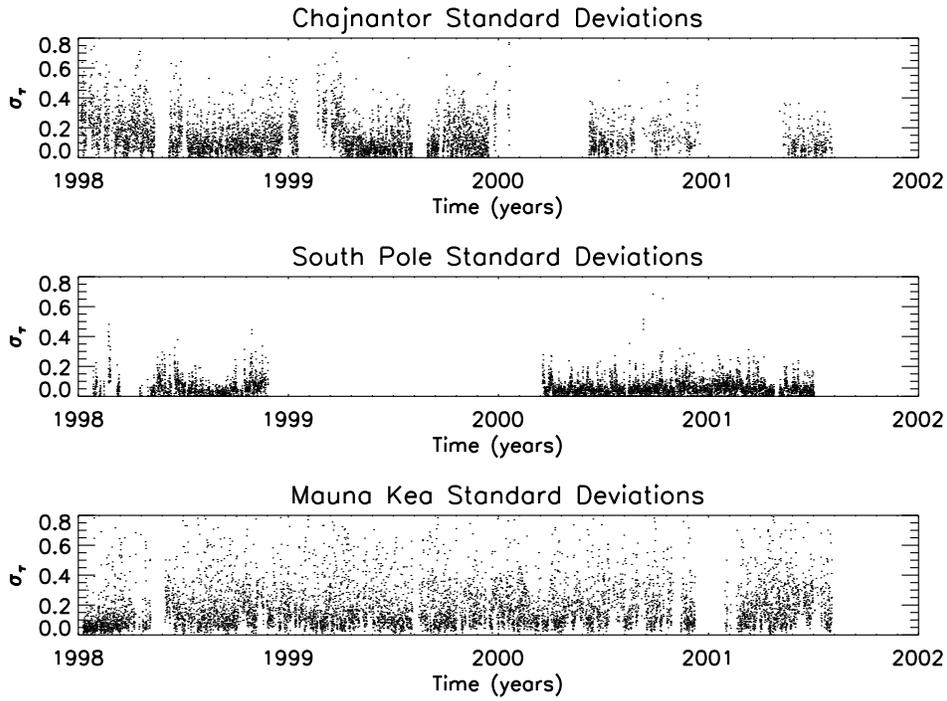}
}
\figcaption{\label{allstd}{\bf Variations of Optical Depth. } 
The standard deviation of $\tau$ during two hour intervals is plotted. Instrument noise has been
subtracted from the measured variance to produce the values plotted.}
\end{figure*}

\begin{figure*}
{\centering\leavevmode\epsfysize=3.75in \epsfbox{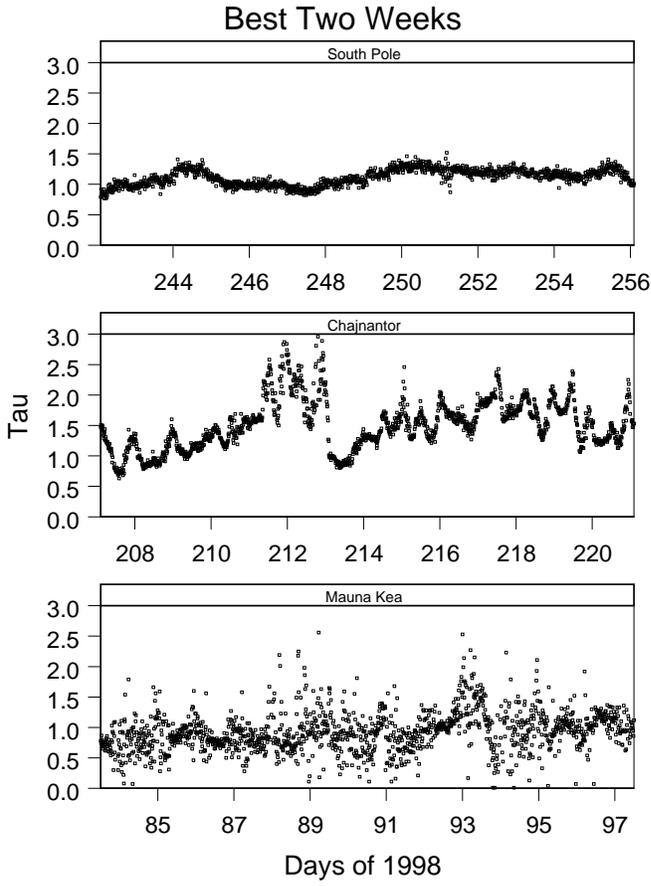}}
\figcaption{\label{allweeks}{\bf Best Two Weeks.\ \ } 
Zenith optical depths at 350 ${\rm \mu m}$ are shown for the most stable two weeks during
1998 at each site.  
The point to point variations in the South Pole data are due to instrumental noise.  
The tippers at the other sites have similar noise levels.  
Strong diurnal variation is evident in the Chajnantor data.  
Mauna Kea has only occasional short periods of stability.}
\end{figure*}

Figure \ref{allyears} shows two hour means of optical depth for the three sites 
during the four years beginning on January 1, 1998. 
Figure \ref{allstd} shows the standard deviation of $\tau$ in two hour intervals for the same period.
Figure \ref{allweeks} 
shows the most stable two week 
period in 1998 for each site.  Successive points are typically taken twelve to fifteen minutes apart.  

\begin{figure*}
{\centering\leavevmode\epsfysize=3.75in 
\epsfbox{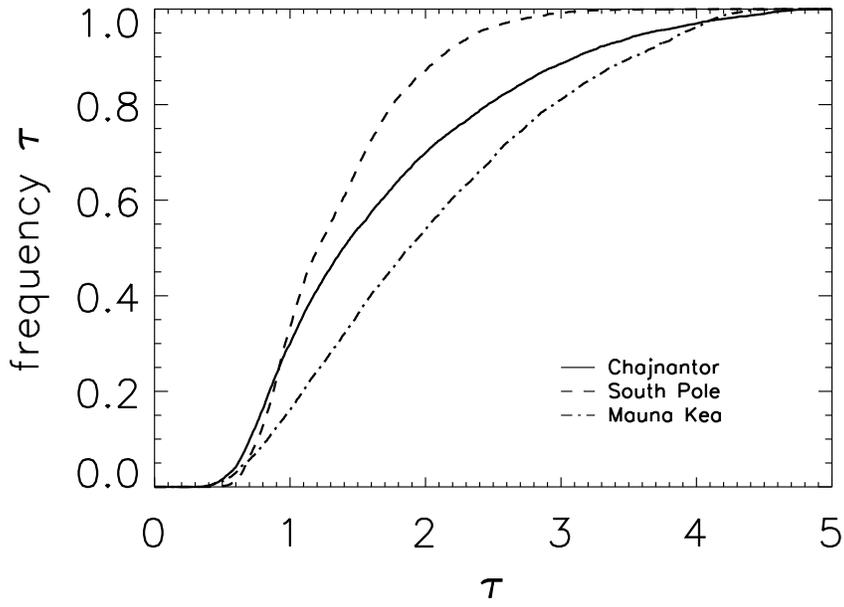}
}
\figcaption{\label{taucdf}{\bf Cumulative Distributions of 860 GHz Zenith Optical Depth.\ \ } 
}
\end{figure*}

The cumulative distributions of optical depth in 1998 are
shown in Figure~\ref{taucdf}, and quartiles from that distribution are shown in table
\ref{taucdftable}.  All three sites have median optical depths between 1 and 2.
Because the transparency of a slab is ${\rm exp}(-A_i \tau)$, 
only 24\% of the radiation from a source at 45 $^\circ$ elevation reaches the telescope when $\tau = 1$. 
These measurements demonstrate the challenge of submillimeter astronomy, even from superb sites.
Below the median the South Pole and Chajnantor have similar distributions of
opacity, while Mauna Kea has higher opacity.

\begin{table}[!hb]
\begin{center}
\begin{tabular}{l|c|c|c}
$\tau$(860 GHz) & 25\% & 50\% & 75\% \\
\hline
South Pole & 0.91 & 1.20 & 1.64 \\
Chajnantor & 0.91 & 1.39 & 2.22 \\
Mauna Kea & 1.20 & 1.88 & 2.73 \\
\end{tabular}
\caption{Quartiles of measured atmospheric zenith optical depth.}
\label{taucdftable}
\end{center}
\end{table}

Our measurements demonstrate the atmosphere at the South Pole is more stable than at the other two 
sites (figures \ref{allyears} and \ref{allstd}). This is particularly evident in figure \ref {allweeks}. 
At the South Pole, variations due to noise intrinsic to the tipper usually exceed real variations of
atmospheric optical depth for averages over periods less than a few hours.
For Chajnantor the sequence of
$\tau$ values follows monotonically increasing or decreasing
trends for periods of several hours. The point to point variations in the Chajnantor data
are mostly due to real changes of optical depth rather than instrumental noise.
Much of this variation is diurnal. Mauna Kea has hardly any 
periods of stability comparable to the other sites.

\section{Quantitative Stability Analysis}
To quantitatively compare the atmospheric stability at the three sites, we studied the fluctuations in
optical depth during two hour intervals.  This choice of interval, 
during which nine observations were typically made, represents a compromise.
Integration periods longer than an hour are needed to average out instrumental noise
sufficiently so variations of optical depth can be measured in the South Pole data.
Integrations longer than a few hours, on the other hand, would be dominated by the diurnal 
variations present at Chajnantor. Also, submillimeter observations tend to last a few hours.

We determined the instrumental noise using the calibration data.
During calibration, the detector voltage, measured while the mirror is
turned toward the warm 
or hot calibrator, is compared with the calibrator temperature, measured using
a thermometer attached to the calibrator.  
To determine the instrumental noise, we calculated the variance
of the detector voltage. The instrument noise level is typically $\sigma_\tau \sim 0.04$

To test whether the noise in the South Pole data, during quite periods, is due to the instrument,
we calculated the set of differences of successive observations of $\tau$ for the South Pole 
during the best two weeks of 1998. 
The resulting distribution of difference is an excellent fit to the Gaussian distribution expected from 
instrumental noise alone.  
Atmospheric variations of optical depth would 
have a broader distribution with long tails. We conclude instrumental noise is dominant 
during the most stable periods and we proceed to subtract its variance from the total variance
to estimate the variance due to atmospheric fluctuations.
Because the South pole is so stable, the variance difference is
negative about 40\% of the time. In these cases the atmospheric varaition is too small to
measure precisely with our instrumental noise and we assign a value of $\sigma_\tau$ in the range 0 to 0.01
to these samples.  

\begin{figure*}
{\centering\leavevmode\epsfysize=3.75in \epsfbox{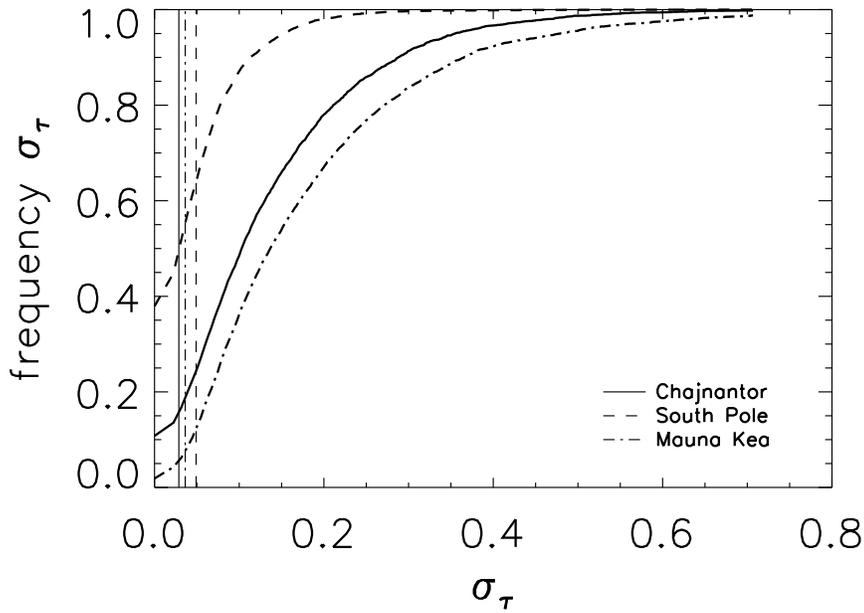}}
\figcaption{\label{comparecorrcdf}
{\bf Cumulative Distribution of 860 GHz Zenith Optical Depth Variations.\ \ } 
These distributions are the 
result of subtraction in quadrature of instrumental noise variance
from the total measured variance. Negative resultant variances are accumulated in the
bin at zero standard deviation.  
We estimate that these variation estimates should be reliable for $\sigma_\tau$ values greater than 0.01.
The vertical lines show the instrumental noise that has been
subtracted. }
\end{figure*}

Cumulative distributions of optical depth variation are
shown in Table \ref{sigmacdftab} and plotted in Figure \ref{comparecorrcdf}.

\begin{figure*}
{\centering\leavevmode\epsfysize=3.75in \epsfbox{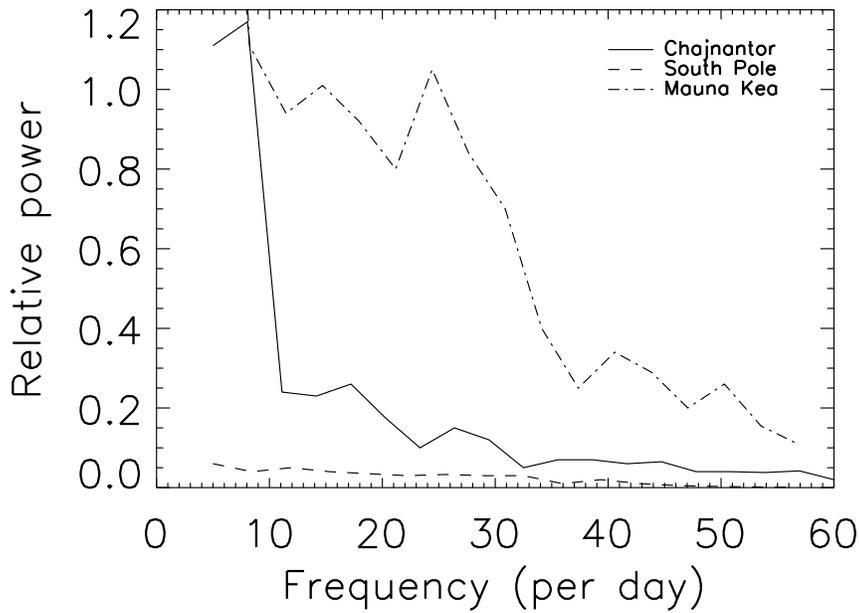}}
\figcaption{\label{fft}{\bf Binned Power Spectra of Zenith Optical Depth.\ \ } 
The irregularly sampled data for each site's best two weeks were converted to a regularly
sampled stream by linear interpolation and the spectrum was calculated. 
The data were binned in frequency and medians
of each frequency range are shown.  
The approximate Nyquist
frequency is 60/day.
The lowest frequency bin of the Mauna Kea power spectrum is off scale with a relative power of 2.53 
at 5/day }
\end{figure*}

We also calculated the power spectrum of the optical depth variations 
for the best-two-weeks-of-1998 segments (Figure \ref{fft}).
The power spectrum falls with frequency at all three sites. The level of optical depth fluctuation power
is much lower at the South Pole than at the other sites.

\begin{table}[!hb]
\begin{center}
\begin{tabular}{l|c|c|c}
$\sigma(\tau)$ & 25\% & 50\% & 75\% \\
\hline
South Pole & $<$0.01 & 0.03 & 0.07 \\
Chajnantor & 0.05 & 0.10 & 0.18 \\
Mauna Kea & 0.08 & 0.13 & 0.24 \\
\end{tabular}
\caption{Quartiles of the sky noise distributions after removal of instrumental noise.}
\label{sigmacdftab}
\end{center}
\end{table}

\section{Conclusion}

At the median, the standard deviation of the zenith optical depth at the South Pole is three
times lower than at Chajnantor and four times lower than at Mauna Kea.
In agreement with Lay \& Halverson (1999), we find that the submillimeter atmospheric opacity
is substantially more stable than at Chajnantor. 
We also find that Chajnantor has better stability 
than Mauna Kea.

\newpage
\section{Acknowledgments}

We thank  South Pole winter-overs Matt Rumitz, Xiaolei Zhang, Matt Newcomb,
Greg Griffin, Roopesh Ojha and Chris Martin and Wilfred Walsh. Keching Xiao did
the added-window experiments at the South Pole. We owe a special debt to Rodney Marks
who died at the South Pole in 2000. This project was supported by NSF grants
AST-9225007 and OPP-8920223 (Center for Astrophysical Research in Antarctica) and
by Carnegie Mellon University.
AST/RO, JCMT and CSO provided space, power and Internet connections for
tippers. Tim Robertson helped design and test an early prototype at the South Pole.
Nigel Atkins, in Hawaii, and Geraldo Valladeres and Angel Ot\'arola, in Chile, assisted with deployments.
ES acknowledges support from the NSF Research Experiences for Undergraduates program at the NRAO.
The National Radio Astronomy Observatory is a facility of the National Science Foundation operated under
cooperative agreement by Associated Universities, Inc.

%
%
%
%

{
\protect\itemsep=0pt
\protect\parsep=0.1\baselineskip
\bibliographystyle{aas}
}
{}

\clearpage

\newpage

\newpage

\newpage

\newpage

\newpage

\end{document}